\documentclass[
showkeys,
twocolumn,
]{revtex4-2}

\usepackage{graphicx}

\begin{document}

\preprint{}

\title{\textbf{Quantum Computing and AI: Perspectives on Advanced Automation in Science and Engineering}}

\author{Tadashi Kadowaki$^{1,2}$}
\affiliation{$^{1}$Global R\&D Center for Business by Quantum-AI Technology, National Institute of Advanced Industrial Science and Technology, Ibaraki, Japan\\$^{2}$DENSO CORPORATION, Tokyo, Japan} 

\begin{abstract}
Recent advances in artificial intelligence (AI) and quantum computing are accelerating automation in scientific and engineering processes, fundamentally reshaping research methodologies. This perspective highlights parallels between scientific automation and established Computer-Aided Engineering (CAE) practices, introducing Quantum CAE as a framework that leverages quantum algorithms for simulation, optimization, and machine learning within engineering design. Practical implementations of Quantum CAE are illustrated through case studies for combinatorial optimization problems. Further discussions include advancements toward higher automation levels, highlighting the critical role of specialized AI agents proficient in quantum algorithm design. The integration of quantum computing with AI raises significant questions about the collaborative dynamics among human scientists and engineers, AI systems, and quantum computational resources, underscoring a transformative future for automated discovery and innovation.
\end{abstract}


\maketitle

%\tableofcontents

\section{Introduction}

Machine learning traces its origins to the perceptron introduced by Rosenblatt in 1958, marking the inception of its development within artificial intelligence \cite{Rosenblatt1958}. Initially constrained by the simplicity of linear classifiers, progress stagnated until breakthroughs in multilayer neural networks and the popularization of the backpropagation algorithm in the late 1980s \cite{Rumelhart1986}. Subsequent innovations, including support vector machines (SVM) by Cortes and Vapnik \cite{Cortes1995}, decision trees \cite{Quinlan1986}, and ensemble methods such as random forests \cite{Breiman2001} and gradient boosting \cite{Friedman2001}, significantly enhanced performance in practical applications.

The 2010s witnessed an explosive growth in deep learning, driven by massive datasets and increased computational capabilities provided by GPUs, which led to outstanding successes in image and speech recognition tasks \cite{LeCun2015}. Architectures like convolutional neural networks (CNNs) demonstrated by Krizhevsky et al. \cite{Krizhevsky2012}, recurrent neural networks (RNNs) by Hochreiter and Schmidhuber \cite{Hochreiter1997}, and Transformers by Vaswani et al. \cite{Vaswani2017} have substantially outperformed previous methods. Particularly in natural language processing (NLP), Transformers facilitated major advancements exemplified by large language models such as GPT \cite{Brown2020} and BERT \cite{Devlin2019}. Utilizing self-supervised learning and enormous text datasets, these models excel in text generation, dialogue systems, and other diverse applications, reshaping the landscape of AI research.

Additionally, advancements in deep reinforcement learning enabled AI to surpass human experts in highly complex games like Go \cite{Silver2016}. Despite its enormous combinational complexity, Go is a complete-information game. Current AI technologies now surpass top human players in even more complex scenarios, such as competitive racing games \cite{Wurman2022}.

Practical industrial applications of AI include advanced driver-assistance systems that enhance automotive safety through onboard cameras and sophisticated feedback mechanisms. Research and development in self-driving technology actively involve automotive manufacturers and IT companies alike, integrating cutting-edge AI techniques, including deep reinforcement learning and large language models.

The ability of AI to exceed human capabilities in specialized tasks suggests transformative potential across scientific and industrial domains. Recognizing this significance, organizations such as the Japan Science and Technology Agency (JST), the National Academies of Sciences (NAS), and the Organisation for Economic Co-operation and Development (OECD) have released strategic proposals concerning AI's integration into scientific discovery and research methodologies \cite{CenterforResearchandDevelopmentStrategy2021,NationalAcademiesofSciences2022,OECD2023}. These initiatives, labeled as ``Novel Turing Challenge,'' ``Accelerated Discovery,'' or ``AI for Science (AI4Science),'' emphasize automating scientific discovery processes, wherein humans define objectives, and AI autonomously formulates hypotheses, performs experiments, and iteratively verifies outcomes.

The notion of scientific automation questions which research activities can be delegated to AI and robotics and explores the resulting efficiencies. AI endows systems with sensory and cognitive capabilities, whereas robotics provides mechanical movement and physical interaction abilities. This technological convergence could ultimately lead to AI-enabled robots assuming roles traditionally reserved for human scientists. Such possibilities have long been explored in science fiction, exemplified by The Greatest Robot on Earth in Osamu Tezuka's Astro Boy, in which a robot scientist surpasses human scientists in both intellect and capability \cite{Tezuka1964}.

As AI and robotics technologies mature, their real-world application in scientific research is anticipated to accelerate, influencing both technical methodologies and societal interactions. Currently, many scientists and engineers operate under corporate or national interests, driving increased demands for productivity and discovery. Integrating AI and robotics to enhance research efficiency thus becomes increasingly essential. The 2024 Nobel Prize in Chemistry awarded to scientists involved in AlphaFold further underscores this emerging trend \cite{Jumper2021}.

Scientific discovery involves extracting novel insights from an expansive space of hypotheses, analogous to mathematician Paul Erdős’s metaphor of discovering theorems recorded in ``The Book'' \cite{Hoffman1998}. Similar to mining gold, the ease of discovery diminishes over time, necessitating innovation to sustain productivity. As long as scientific discoveries promise economic returns exceeding their investment, research will continue, fostering advances in supporting technologies.

King et al. introduced the concept of robot scientists, demonstrating their effectiveness in functional genomics research \cite{King2004}. Automation leveraging AI and robotics has become prevalent, especially in resource-intensive research areas like genome sequencing and compound library screening.

Similar to the disruptive impact IT had on service sectors, recent AI advancements are poised to extend automation into various scientific fields beyond life sciences, making automation indispensable for scientists and engineers across disciplines and scales.

This paper specifically discusses research activities entirely executable within computational environments for two reasons: firstly, improvements in computational power and simulation algorithms have increased the feasibility of replacing physical experiments; secondly, the practical realization of quantum computing promises to further enhance these capabilities.

Fully automated computational research processes --- spanning topic selection, programming and debugging, conducting computational experiments, data analysis, manuscript drafting, and even peer review --- are already underway \cite{Lu2024}. Subsequent studies have demonstrated instances of AI-authored papers successfully passing peer review, highlighting the potential for AI agents to efficiently handle demanding research tasks within projects.

In this paper, we overview the concept of scientific automation, explore its parallels with widely adopted Computer-Aided Engineering (CAE), introduce strategies and examples of quantum computing applications within CAE, discuss advancements towards higher automation levels, and conclude with future perspectives.

\section{Automation in Science and Engineering}

As generative AI approaches human-like capabilities, the concept of ``digital humans'' emerges. Analogously, AI systems conducting research and development through computer simulations can be termed ``digital scientists.'' Unlike digital humans, digital scientists do not require communication skills. Instead, they must possess the capability for rapid logical decision-making grounded in scientific knowledge. Consequently, digital scientists neither require distinct personalities nor can autonomously undertake undesired research beyond the scope of human oversight.

Using the term ``digital scientist'' strategically positions AI as a supportive tool rather than a replacement for human scientists, thereby mitigating social concerns. The automation of scientific research thus fosters collaborative frameworks between human scientists and digital scientists.

But what capabilities must digital scientists possess? The strategic report from JST categorizes scientific automation into six levels (from Level 0 to Level 5), analogous to the autonomy levels for self-driving vehicles defined by SAE International. Level 3 automation corresponds to full autonomy under well-defined conditions, often referred to as ``perfect information.''

The scientific discovery process, illustrated by the work of King et al., typically involves (1) hypothesis generation based on existing knowledge, (2) experimental design and execution to verify these hypotheses, and (3) data analysis to validate hypotheses. This cyclic process continuously updates knowledge and generates new hypotheses, constituting scientific automation.

At Level 3 in self-driving technology, human oversight ensures AI properly manages steering, acceleration, and braking, intervening only when necessary. Similarly, scientific automation requires human scientists to provide sufficient information and evaluation criteria to AI, automating routine tasks while humans interpret the purpose and implications of results.

Levels 4 and 5 automation require complete adaptability to dynamically changing conditions, capabilities currently beyond practical realization. Analogously, in science, comprehensive automation remains in foundational research stages. Nevertheless, pioneering studies aim at fully automated scientific processes from topic selection to manuscript preparation, as described in the previous section.

King et al. demonstrated Level 3 scientific automation in life sciences, a field inherently reliant on experiments. Conversely, disciplines like materials science employ first-principles calculations enabling complete computational workflow, broadly facilitating scientific automation.

This automated methodology closely parallels Computer-Aided Engineering (CAE), widely employed in manufacturing and product design. CAE employs numerical techniques --- such as the finite element method (FEM), boundary element method (BEM), and computational fluid dynamics (CFD) --- to predict product characteristics based on design parameters. Widely adopted across automotive, aerospace, electronics, material industries, CAE optimizes designs, reduces prototype costs, shortens development cycles, ensures quality control, and supports digital transformation (DX). Here, automation in product design will be termed ``design automation.''

In design automation, numerical parameters termed ``design variables $(x)$'' are simulated to determine their relationship with ``product characteristics $(y)$.'' Machine learning then models these characteristics from data, creating predictive models analogous to scientific knowledge. Solving inverse problems identifies optimal values of design variables $(x^*)$ that yield desired (maximized or minimized) product characteristics $(y^*)$. Iteratively refining this optimization process drives design automation forward.

\begin{figure}[t]
  \centering
  \includegraphics[width=8cm]{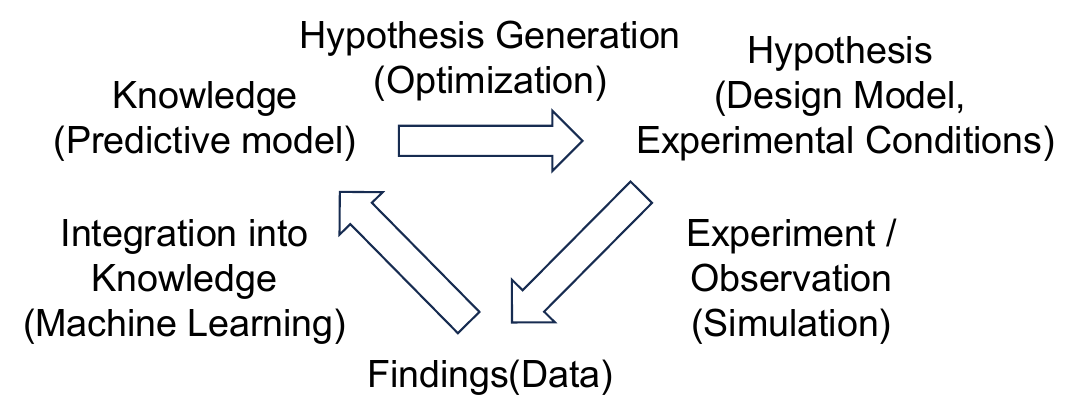}
  \caption{Process of automation in science and product design. Hypotheses are automatically generated based on existing knowledge and validated through experiments or simulations, producing new findings. These findings are integrated into the knowledge base, enabling iterative hypothesis generation. Parentheses indicate specific information processing tasks involved in the automation process. This process parallels design automation workflows such as those found in Computer-Aided Engineering (CAE).}
  \label{fig1}
\end{figure}

Comparatively, scientific automation parallels design automation: (1) Optimization corresponds directly to hypothesis generation, (2) simulation parallels experimental verification, and (3) machine learning aligns with the analysis of resulting data. Initially limited data yields low model accuracy, which progressively improves with iterative simulations and model updates. Figure \ref{fig1} illustrates this cyclic workflow.

This optimization method, ``black-box optimization,'' addresses unknown input-output relations, with Bayesian optimization often applied for continuous variables. Conversely, combinatorial optimization problems involving discrete variables pose greater challenges and have primarily relied on relaxation techniques until recently.

The subsequent chapters will explore quantum computing applications within CAE and examples of their implementation, advancing the discourse on automation in science and engineering.

\section{Quantum CAE}

Quantum computing presents significant opportunities for enhancing CAE by leveraging quantum algorithms \cite{Dalzell2025} in simulation, machine learning, and optimization, which together are termed Quantum CAE. Integrating quantum computing into product design automation could significantly reduce development lead times, resulting in improved productivity, cost-efficiency, and market responsiveness.

Implementing Quantum CAE requires the development of quantum algorithms tailored to typical CAE simulations, as well as quantum machine learning models and quantum optimization methods to process simulation results efficiently. Quantum optimization is particularly promising for handling discrete design variables commonly encountered in engineering design. Furthermore, seamlessly integrating these quantum tasks through coherent quantum information processing technologies is critical.

Initial stages of Quantum CAE involve executing one of the key tasks (simulation, machine learning, or optimization) using quantum computers. For example, employing quantum annealing for optimization tasks to identify candidate designs (hypotheses) represents a practical starting point. Leveraging existing technologies such as quantum gate simulators or Ising solvers can demonstrate short-term, practical benefits on small-scale problems.

Quantum-inspired classical algorithms, such as those developed by Tang derived from quantum recommendation techniques, demonstrate efficient performance on conventional computers, illustrating how quantum research advances classical computing \cite{Tang2019}.

The long-term goal of Quantum CAE is fully quantum-integrated CAE processes. However, significant challenges persist, especially the exponential computational cost associated with encoding classical data into quantum states and extracting useful information back into classical form in the worst case. Overcoming these limitations necessitates methods facilitating efficient quantum state information exchange among tasks. Current research efforts focus on facilitating direct quantum-state exchanges between quantum simulations and quantum machine learning models, and developing quantum machine learning techniques capable of learning from limited quantum data \cite{Huang2022}. Additionally, methods integrating quantum annealing outputs with quantum gate processing to improve optimization effectiveness have been demonstrated \cite{Kadowaki2024}.

Quantum CAE algorithms fundamentally solve complex optimization problems, aligned with general quantum search algorithms such as Grover adaptive search \cite{Gilliam2021} and quantum walks \cite{Marsh2020}. Unlike abstract quantum oracle-based algorithms, Quantum CAE emphasizes direct integration of concrete physical simulations and evaluation of product characteristics, creating practical, efficient quantum circuits suitable for real-world applications. Such developments highlight the practical realization of foundational quantum computing research, potentially driving entirely new algorithmic approaches.

Quantum CAE at Level 3 automation can address highly complex design challenges, including optimizing fusion reactor designs through advanced plasma simulations and comprehensive design optimizations for interstellar spacecraft. Higher automation levels (Levels 4 and 5) will require tackling complex, uncontrollable societal challenges and investigating scientific frontiers beyond current human cognitive capacities. Quantum computing could become indispensable, propelled by Quantum CAE-derived knowledge and technologies.

\section{Quantum CAE for Manufacturing}

This chapter presents practical examples of Quantum CAE applications at Level 3 automation, focusing specifically on discrete-variable design problems solved using black-box optimization methods.

Baptista and Poloczek introduced Bayesian Optimization of Combinatorial Structures (BOCS), a method that constructs probabilistic prediction models by sampling from the posterior distribution over model parameters \cite{Baptista2018}. Kitai et al. proposed Factorization Machine Quantum Annealing (FMQA), employing Factorization Machines (FM) as prediction models and quantum annealing for optimization \cite{Kitai2020}. FM differs from Gaussian process methods used in continuous optimization by employing point estimation techniques, enabling faster convergence with fewer data points but increasing the risk of convergence to local optima. Conversely, BOCS utilizes distribution-based modeling similar to Gaussian processes, facilitating broader exploratory searches.

These methods rely on discrete design variables and evaluation approaches, such as simulations, to assess product characteristics. Design optimization problems involving discrete variables are typically NP-hard, complicating the task of obtaining high-quality solutions. Both BOCS and FMQA efficiently derive feasible solutions from limited iterations.

Two case studies illustrate the practical application of these methods:

First, Matsumori et al. addressed mounting-point optimization in automotive electronic control boards \cite{Matsumori2022}. To mitigate vibration-related risks in automotive engine rooms, the board’s lowest natural frequency must be maximized to avoid resonance and minimize displacement. Increasing the lowest natural frequency typically necessitates additional mounting points, creating trade-offs between improved stability, design flexibility, and manufacturing cost. Formulating this as a multi-objective optimization problem, BOCS and FMQA efficiently identified optimal number of mounting points and their locations.

\begin{figure}[t]
  \centering
  \includegraphics[width=4cm]{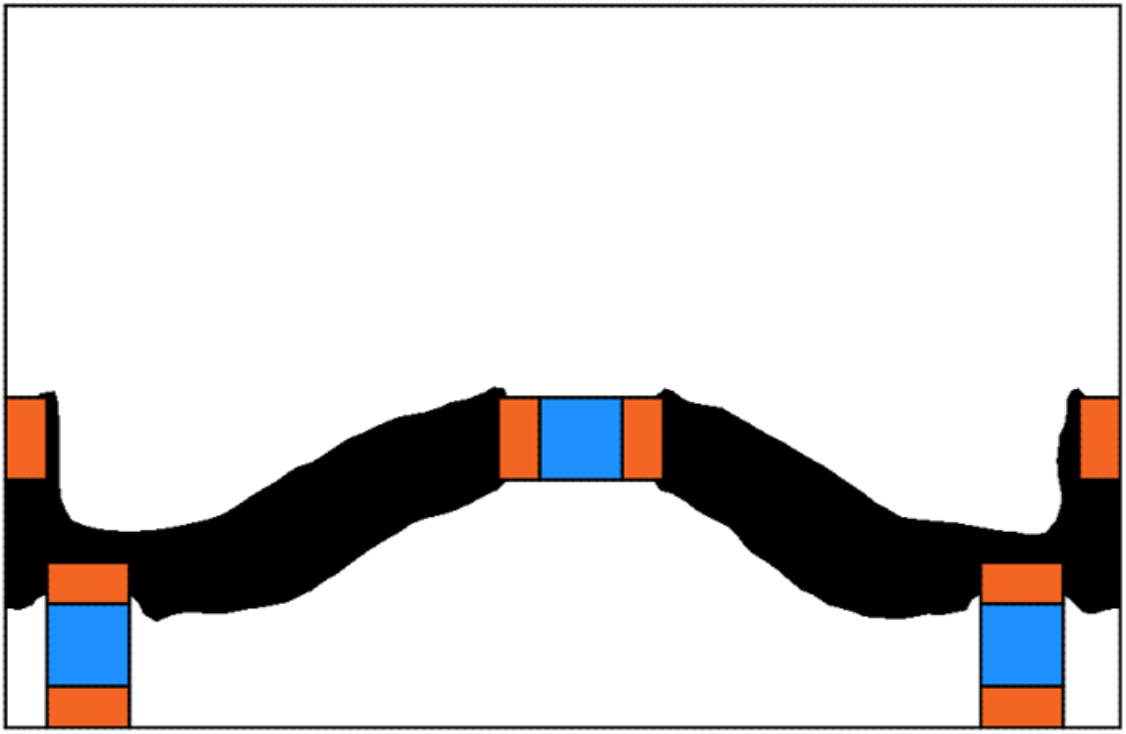}
  \includegraphics[width=4cm]{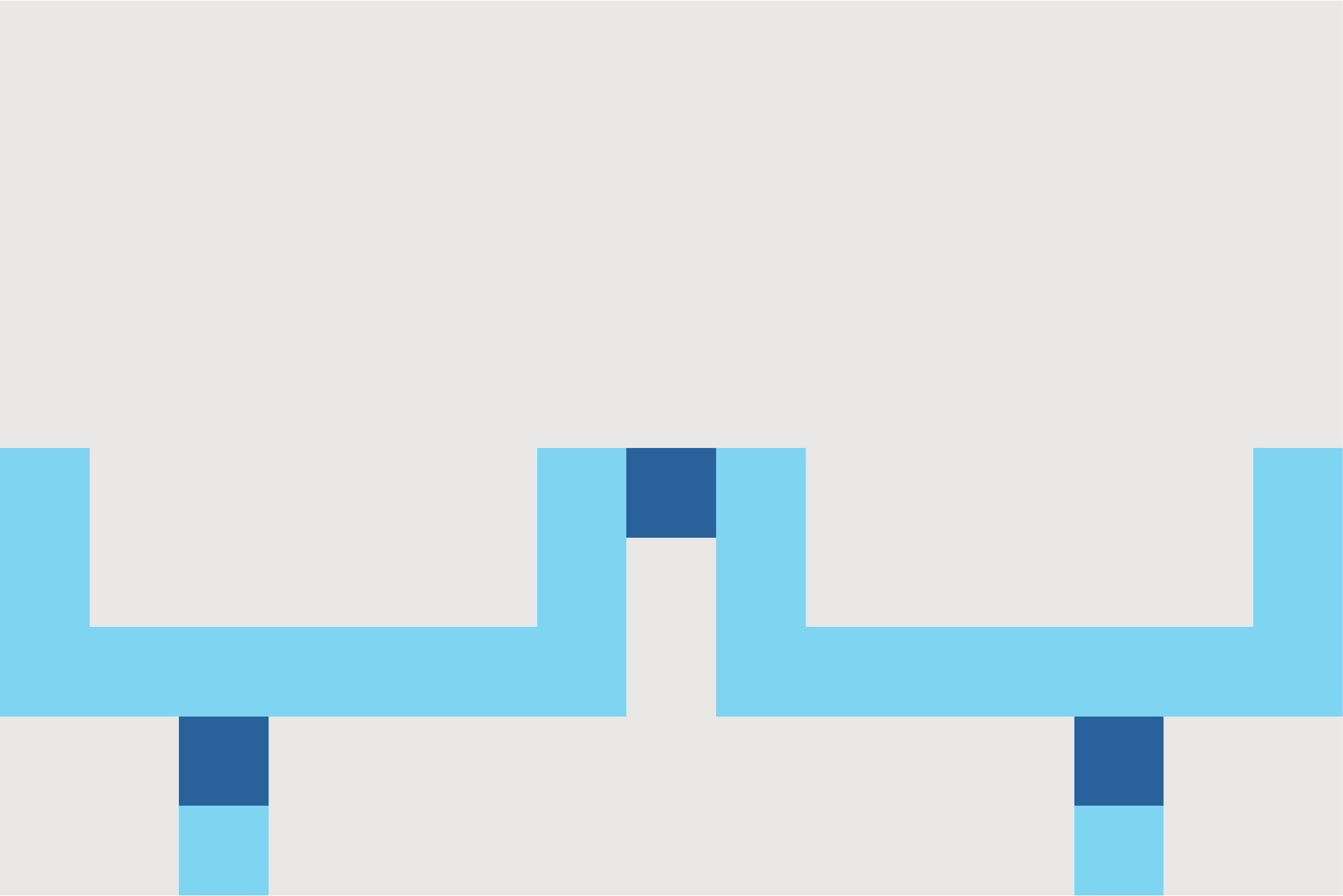}
  \caption{Optimization results for a noise filter. Similar solutions were obtained using topology optimization (left, reproduced from \cite{Maruyama2021} with permission) and black-box optimization (right, reproduced from \cite{Okada2023}).}
  \label{fig2}
\end{figure}

\begin{figure}[t]
  \centering
  \includegraphics[width=8cm]{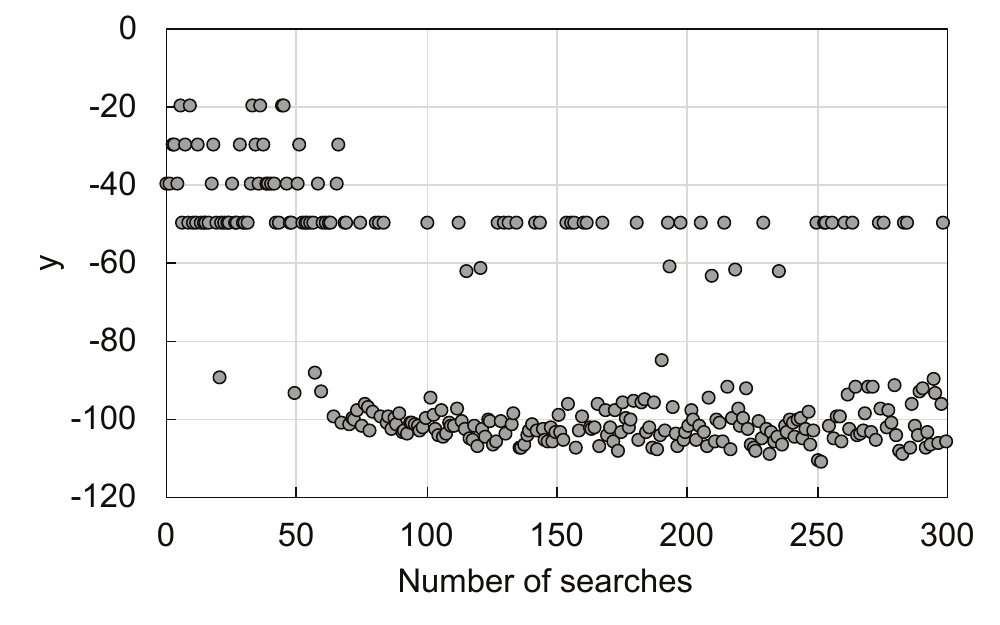}
  \caption{Relationship between the number of data acquisitions (x-axis) and the cost function (y-axis). Feasible solutions emerge as the number of data acquisitions increases. (reproduced from \cite{Okada2023})}
  \label{fig3}
\end{figure}

Next, Okada et al. considered the design of high-frequency noise filters \cite{Okada2023}. High-frequency circuits require evaluation as distributed-element circuits, accounting for interactions between printed circuit patterns and electronic components. Traditional rule-based methods, AI-based approaches, and topology optimization \cite{Maruyama2021} techniques have been previously applied. Multiple candidates of printed circuit patterns between components were prepared, formulating the design as a combinatorial optimization problem. A cost function employing constraints aligned with Quadratic Unconstrained Binary Optimization (QUBO) was developed to efficiently exclude impractical printed circuit patterns. As a result, practical and optimal printed circuit patterns were effectively identified. Figure \ref{fig2} shows that the solutions obtained by topology optimization and black-box optimization methods are essentially similar. Figure \ref{fig3} demonstrates that iterative data acquisition leads to an exploration shift from impractical to practical printed circuit patterns.

Beyond these case studies, potential applications have been explored in various fields, including pharmaceuticals \cite{Khan2023,Tus2023,Mao2023}, chemistry \cite{Gao2023}, optics \cite{Kim2022,Zhu2022,Inoue2022}, magnetic materials \cite{Nawa2023,Maruo2022}, electronics \cite{Dou2022,Oh2022}, and data science \cite{Drouet2020,Kadowaki2022}. These case studies highlight the effectiveness of quantum annealing as well as Ising solvers in enhancing automation frameworks in both manufacturing and scientific fields. If larger-scale quantum annealers become available, prediction model construction could become a bottleneck. Therefore, while quantum algorithms represent a long-term research goal, improving conventional computing algorithms remains an important short-term objective \cite{Kadowaki2022,Minamoto2025}.

Furthermore, Quantum CAE realization also demands quantum algorithm implementations for simulations. Inspired by quantum algorithms for the Boltzmann equation \cite{Budinski2021}, Igarashi et al. developed a quantum algorithm for solving radiative transfer equations, a class of linear differential equations, which allows efficient implementation via quantum circuits \cite{Igarashi2024}. Nonlinear equations, frequently encountered in real-world simulations, require tailored quantum algorithm development \cite{Liu2021,Joseph2020}. Additionally, research into implementations using quantum circuits \cite{Layden2023} and quantum annealers \cite{Arai2025} to generate proposal states with reduced rejection rates for accelerating Monte Carlo simulations exemplifies ongoing advancements in quantum-enhanced computational methods.

\section{Perspectives for Advanced Automation}

The previous chapters focused primarily on Level 3 automation within scientific and engineering contexts. Advancing to higher automation levels (Levels 4 and 5) necessitates developing sophisticated teams of AI agents capable of autonomously conducting comprehensive research activities --- from formulating research topics and conducting experiments to data analysis and manuscript writing. Achieving this goal requires enhancing individual agent capabilities and developing advanced technologies for effective collaboration and communication among agents within an integrated framework.

Developing practical AI agents requires a combination of versatile, general-purpose agents and highly specialized agents equipped with expert knowledge or specialized skills. Quantum computing's role in accelerating scientific and design processes underscores the importance of developing specialized agents proficient in quantum algorithms and quantum circuit design. Recent research has explored generative models that autonomously design quantum circuits tailored to specific objectives. For instance, Nakaji et al. introduced the Generative Quantum Eigensolver (GQE), employing a decoder-only Transformer to compute molecular ground states through automated quantum circuit generation \cite{Nakaji2024}. Extending this concept further, Minami et al. introduced an AI agent capable of autonomously generating quantum circuits specifically designed for solving optimization problems \cite{Minami2025}.

\begin{figure}[t]
  \centering
  \includegraphics[width=8cm]{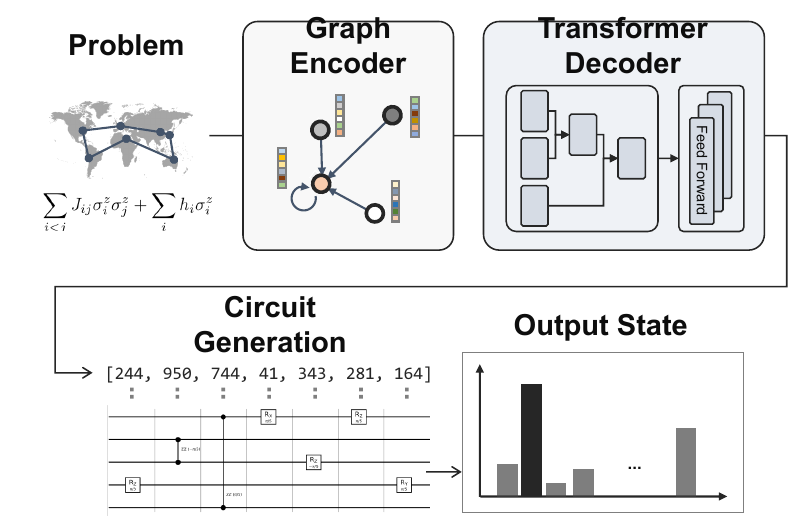}
  \caption{Encoder-decoder model for generating quantum circuits to solve optimization problems. The QUBO matrix, which encodes an optimization problem, is first converted into a graph and then processed using a graph neural network as the encoder and a transformer as the decoder. (reproduced from \cite{Minami2025} with modifications)}
  \label{fig4}
\end{figure}

Discrete optimization problems are commonly formulated using Quadratic Unconstrained Binary Optimization (QUBO). Since QUBO matrices generalize adjacency matrices used in graph theory, these problems naturally lend themselves to graphical representations. Capitalizing on this graphical representation, an AI agent employing an encoder-decoder neural network architecture was developed. Specifically, the encoder uses a Graph Neural Network (GNN) to process input graph data, while the decoder employs a Transformer to generate quantum circuits for solving optimization problems. Performance assessments using quantum circuit simulators guided reinforcement learning strategies, demonstrating the AI agent’s effectiveness in generating quantum circuits (Fig. \ref{fig4}).

Sakka et al. provide a practical example of an AI scientist applying automated quantum machine learning for feature map design by generating ideas, implementing and validating code, and analyzing results, thereby demonstrating the potential of AI-driven quantum algorithm development in practical settings \cite{Sakka2025}.

Developing specialized agents for quantum computing increasingly demands interdisciplinary collaboration between quantum computing specialists and AI experts, alongside integrating advancements from broader AI research. Quantum computing and AI are both developing technologies, and advancements in each independently contribute to automating scientific and engineering processes. Additionally, AI applications are actively being employed to facilitate advancements in quantum computing itself \cite{Alexeev2024}.

\section{Summary and Future Perspectives}

Rapid advancements in AI technology have significantly accelerated automation across various scientific and engineering domains. This paper highlighted how current technological capabilities enable Level 3 automation, comparable to widely established CAE practices in industry. Furthermore, integrating quantum computing into these automated processes offers the potential to significantly enhance their efficiency. Specific examples were presented illustrating how Quantum CAE can deliver tangible benefits even at small scales.

Future research should prioritize two main strategies: First, practical application of existing quantum technologies to specific product development and engineering problems. Second, further development of advanced quantum algorithms in anticipation of future improvements in quantum computing capabilities. Insights and technologies gained through Quantum CAE will not only facilitate advanced scientific research automation but also expand the role of quantum computing toward achieving higher automation levels (Levels 4 and 5).

Central to this vision is the development of advanced AI agents, or ``digital scientists,'' designed to handle specialized tasks and actively collaborate with human scientists. To advance this direction, it is essential to establish integrated infrastructure supporting joint quantum computing and AI research. The current adoption of quantum computing at supercomputing centers worldwide presents an opportunity to build robust, integrated research communities focused on quantum computing and AI.

Lastly, we are approaching an era in which AI surpasses human capabilities in various research domains. Concurrently, AI is gaining access to computational resources that provide quantum acceleration. However, fundamental questions remain regarding how scientists and engineers, businesses, nations, and society will establish effective collaboration frameworks with AI. Answers to these questions will progressively emerge through continuous dialogue between humans and AI systems. Such interactions will leverage AI to efficiently navigate humanity’s extensive knowledge base, stimulate innovative discoveries, foster deeper intellectual engagement, and support ethical approaches to scientific and technological development.

\section*{Acknowledgments}

The author gratefully acknowledges Mitsuru Ambai, Tadayoshi Matsumori, Masato Taki, Akihisa Okada, Asuka Igarashi, Shiro Kawabata, Shunya Minami, Kohei Nakaji, Yoichi Suzuki, Alán Aspuru-Guzuk, Shunta Arai, and Tatsuya Ishigaki for fruitful collaboration and insightful discussions in joint research activities. This work was partly performed for Council for Science, Technology and Innovation (CSTI), Cross-ministerial Strategic Innovation Promotion Program (SIP), ``Promoting the application of advanced quantum technology platforms to social issues''(Funding agency: QST). This paper was partly based on results obtained from a project, JPNP16007, commissioned by the New Energy and Industrial Technology Development Organization (NEDO), Japan.

%\bibliography{main}

\end{document}